\documentclass[a4paper,11pt]{article}
\usepackage{pos}
\usepackage{float}
\usepackage[rflt]{floatflt}

\newcommand{\GeV}{{{\sf GeV}} }

\title{
The variable flavor number scheme to three-loop
order\footnote{DESY 26--064, RISC Report number 26-06, CERN-TH-2026-113, MPP-2026-89,
PoS (LL2026) 025
}}
\ShortTitle{The VFNS to three-loop order}

\author[a]{J.~Ablinger}   
\author[b]{A.~Behring}
\author*[c,d]{J.~Bl\"umlein}
\author[a]{A.~De Freitas}
\author[e]{A.~von Manteuffel}
\author[a]{C.~Schneider}
\author[f]{K.~Sch\"onwald}

\affiliation[a]{Johannes Kepler University Linz, Research Institute for Symbolic
Computation (RISC), Altenberger Stra\ss{}e 69, A-4040, Linz, Austria}

\affiliation[b]{Max-Planck-Institut f\"ur Physik,
Boltzmannstra\ss{}e 8, 85748 Garching, Germany}

\affiliation[c]{Deutsches Elektronen-Synchrotron DESY, Platanenallee 6, 15738 Zeuthen,
Germany}

\affiliation[d]{Institut f\"ur Theoretische Physik III, IV, TU Dortmund, Otto-Hahn
Stra\ss{}e 4, 44227 Dortmund, Germany}

\affiliation[e]{Institut f\"ur Theoretische Physik, Universit\"at Regensburg, 93040
Regensburg, Germany}

\affiliation[f]{CERN, Theoretical Physics Department,
CH-1211 Geneva 23, Switzerland}

\emailAdd{Jakab@gmx.at}
\emailAdd{arnd.behring@desy.de}
\emailAdd{Johannes.Bluemlein@desy.de}
\emailAdd{abilio.de.freitas@desy.de}
\emailAdd{manteuffel@ur.de}
\emailAdd{Carsten.Schneider@risc.jku.at}
\emailAdd{kay.schonwald@cern.ch}

\abstract{We describe the variable flavor number scheme to three-loop order, which modifies 
the massless parton densities by single- and two-mass effects and introduces heavy-quark 
parton distribution functions for charm and bottom. A renormalization group analysis 
shows the validity of this picture at large scales $Q^2$, where it resembles the 
non-power-suppressed heavy-flavor corrections completely. We also provide numerical 
implementations of a series of charged and neutral current Wilson coefficients.}

\FullConference{Loops and Legs in Quantum Field Theory (LL2026)\\
12-17, April, 2026\\
Bayreuth, Germany\\}

\begin{document}
\maketitle

\section{Introduction}
\label{sec:1}

\vspace*{1mm}
\noindent
The variable flavor number schemes (VFNS) account for the universal heavy-flavor 
contributions
of the deep-inelastic (DIS) structure functions in the form of modified parton distribution
functions. They are important for the matching at flavor thresholds. In the region of large
virtualities $Q^2 \gg m_Q^2$, where $m_Q$ denotes the  heavy-quark mass, full account is
given of all heavy-quark corrections, neglecting power corrections of $O((m^2/Q^2)^k),~k 
\geq 1$.
This is implemented by the renormalization group equation \cite{Callan:1970yg,Symanzik:1970rt}.
Since the ratio of the charm and bottom
quark masses is not small, one may consider the decoupling of two quarks at the same time.
The single-mass VFNS has been introduced in Ref.~\cite{Buza:1996wv} to two-loop order and 
two-loop two-mass effects were described in Refs.~\cite{Blumlein:2018jfm,Bierenbaum:2022biv}.
The single-mass VFNS at three-loop order has been presented in 
Ref.~\cite{Ablinger:2025joi,Behring:2025avs}
in the unpolarized and polarized cases. The structure of the two-mass VFNS to three-loop 
orders
was described in Ref.~\cite{Ablinger:2017err}. Using the VFNS, massless Wilson coefficients are 
paired with massive operator matrix elements (OMEs). This does not allow the description of 
power corrections arising in observables, which will become important in the kinematic 
region of
virtualities $Q^2$ below a certain scale $Q_0^2$. For the structure function $F_2(x,Q^2)$
one has $Q_0^2 = 10~m_Q^2$, cf.~\cite{Buza:1995ie}. 

Some modifications of the single-mass VFNS have been discussed in the literature in 
Refs.~\cite{Aivazis:1993pi,Collins:1998rz,Cacciari:1998it,Thorne:1997ga,Forte:2010ta},
to second order. Also these approaches have to be shown to obey the renormalization
group equation to construct observables using the respective modified parton distribution
functions.

This paper is organized as follows. Basic relations of the VFNS are discussed 
in Section~\ref{sec:2}. In Section~\ref{sec:3} the heavy-quark corrections to 
structure functions at large virtualities are described.  Quantitative results on
the single-mass heavy-quark effects on parton distribution functions are described in
Section~\ref{sec:4}. The two-mass operator matrix elements are illustrated in Section~\ref{sec:5}
and Section~\ref{sec:6} contains the conclusions.
\section{Operator matrix elements and the VFNS}
\label{sec:2}

\vspace*{1mm}
\noindent
The massive OMEs, cf.~Ref.~\cite{Bierenbaum:2009mv}, have all been calculated in the 
single- and two-mass cases to three-loop order. 
The one-loop corrections were obtained in Refs.~\cite{Witten:1975bh,Watson:1981ce}, the two-loop 
corrections in Refs.~\cite{Buza:1995ie,Buza:1996wv,Buza:1996xr,Bierenbaum:2007qe,Bierenbaum:2009zt,
Blumlein:2016xcy,Bierenbaum:2022biv}. At three-loop order the single-mass OMEs were 
calculated in
Refs.~\cite{Ablinger:2010ty,Behring:2014eya,Ablinger:2014lka,Ablinger:2014vwa,Ablinger:2014nga,
Ablinger:2019etw,Behring:2021asx,Blumlein:2021xlc,Ablinger:2022wbb,Ablinger:2023ahe,Ablinger:2024xtt}
and in the two-mass case in Refs.~\cite{Ablinger:2017err,Ablinger:2017xml,Ablinger:2018brx,
Ablinger:2019gpu,Ablinger:2020snj,Ablinger:2025nnq}.
At one-loop order only the OMEs $A_{Qg}, A_{gg,Q}$ contribute, at two-loop order 
one further has $A_{qq,Q}^{\rm NS}, A_{Qq}^{\rm PS}, A_{gq,Q}$. Finally, at three-loop order 
there are 
also contributions due to $A_{qq,Q}^{\rm PS},A_{qg,Q}$ in the single-mass case. If the first index
contains only lower case letters, the local operators are placed on massless lines.
The first two-mass contributions emerge at the two-loop level with factorizing contributions
$A_{Qg}^{(1)} \cdot A_{gg,Q}^{(1)}$ \cite{Blumlein:2018jfm,Bierenbaum:2022biv}.
Genuine non-factorizing contributions first emerge at three-loop order
 $A_{qq,Q}^{\rm NS, two},
 A_{Qq}^{\rm PS, two},
 A_{Qg}^{\rm two},
 A_{gq,Q}^{\rm two},
 A_{gg,Q}^{\rm two}$, supplemented by factorizing contributions.

In the polarized case, we work in the Larin scheme \cite{Larin:1993tq}, which requires 
to refer to parton distributions evolved in this scheme, cf.~Ref.~\cite{Blumlein:2024euz}, and 
Wilson coefficients \cite{Blumlein:2022gpp} or hard-scattering sub-system cross sections in this 
scheme to form observables.

In Mellin space the parton densities in the two-flavor VFNS are obtained from the 
massless parton distributions by the following relations,

\vspace*{-1cm}
\begin{eqnarray}
\lefteqn{f_k(N_F+2) + f_{\overline{k}}(N_F+2) =} \nonumber\\ &&
A_{qq,Q}^{\sf NS,+}\left(N_F+2,\frac{m_1^2}{\mu^2},\frac{m_2^2}{\mu^2}\right)
              \cdot \bigl[f_k(N_F)+f_{\overline{k}}(N_F)\bigr]
+ \frac{1}{N_F} A_{qq,Q}^{\sf PS}\left(N_F+2,\frac{m_1^2}{\mu^2},\frac{m_2^2}{\mu^2}\right)
              \cdot\Sigma^+(N_F) \nonumber\\
           &+& \frac{1}{N_F} 
A_{qg,Q}\left(N_F+2,\frac{m_1^2}{\mu^2},\frac{m_2^2}{\mu^2}\right)
              \cdot G(N_F)
    \\
\lefteqn{f_k(N_F+2) - f_{\overline{k}}(N_F+2) = 
{A_{qq,Q}^\text{NS,-}\left(N_F+2,\frac{m_1^2}{\mu^2},\frac{m_2^2}{\mu^2}\right)}
              \cdot\bigl[f_k(N_F)-f_{\overline{k}}(N_F)\bigr]}
    \\
\lefteqn{f_Q(N_F+2) + f_{\overline{Q}}(N_F+2) =} \nonumber\\ &&
{A_{Qq}^\text{PS}\left(N_F+2,\frac{m_1^2}{\mu^2},\frac{m_2^2}{\mu^2},\right)}
              \cdot \Sigma^+(N_F)
           + 
{A_{Qg}\left(N_F+2,\frac{m_1^2}{\mu^2},\frac{m_2^2}{\mu^2}\right)}
              \cdot G(N_F)  \\
\lefteqn{f_c(N_F+1) - f_{\overline{c}}(N_F+1) =
A_{Qq}^\text{PS,s}\left(N_F,\frac{m_c^2}{\mu^2}\right)
              \cdot \Sigma^-(N_F)}\\
\lefteqn{f_b(N_F+1) - f_{\overline{b}}(N_F+1) =
A_{Qq}^\text{PS,s}\left(N_F,\frac{m_b^2}{\mu^2}\right)
              \cdot \Sigma^-(N_F)} \\
\lefteqn{\Sigma^+(N_F+2) =} \nonumber\\ &&
\Biggl[
{A_{qq,Q}^\text{NS}\left(N_F+2,\frac{m_1^2}{\mu^2},\frac{m_2^2}{\mu^2}\right)}
+ {A_{qq,Q}^\text{PS}\left(N_F+2,\frac{m_1^2}{\mu^2},\frac{m_2^2}{\mu^2}\right)}
                + 
{A_{Qq}^\text{PS}\left(N_F+2,\frac{m_1^2}{\mu^2},\frac{m_2^2}{\mu^2}\right)}
              \Biggr] \nonumber\\  
              && \times \Sigma^+(N_F) +
\Biggl[
{A_{qg,Q}\left(N_F+2,\frac{m_1^2}{\mu^2},\frac{m_2^2}{\mu^2}\right)}
+{A_{Qg}\left(N_F+2,\frac{m_1^2}{\mu^2},\frac{m_2^2}{\mu^2}\right)}
              \Biggr]   \cdot G(N_F)
    \\
\lefteqn{G(N_F+2) =} \nonumber\\ 
          &&
A_{gq,Q}\left(N_F+2,\frac{m_1^2}{\mu^2},\frac{m_2^2}{\mu^2}\right)
                        \cdot \Sigma^+(N_F)
+ A_{gg,Q}\left(N_F+2,\frac{m_1^2}{\mu^2},\frac{m_2^2}{\mu^2}\right)
                        \cdot G(N_F)~.
\end{eqnarray}
Here $N_F$ denotes the number of light flavors, $f_k$ are the light-quark densities and 
$G$ denotes the gluon distribution. $A_{qq}^{\rm NS,+}$ refers to the even and $A_{qq}^{\rm NS,-}$
to the odd moments of $A_{qq}^{\rm NS}$ in the unpolarized case.
The singlet distributions $\Sigma^\pm$ are given by
\begin{eqnarray}
         \Sigma^\pm &=& [u \pm \bar{u}] + [d \pm \bar{d}]  + [s \pm \bar{s}].
\end{eqnarray}
\section{Heavy-quark corrections to structure functions at large virtualities}
\label{sec:3}

\vspace*{1mm}
\noindent
In the fixed flavor number scheme the asymptotic heavy-quark corrections
to the structure function $F_2(x,Q^2)$, disregarding 
power corrections, are given by Eq.~(10) of Ref.~\cite{Ablinger:2025awb}. In the 
single-mass VFNS one obtains
\begin{eqnarray}
F_2^{\rm VFNS}(x,Q^2) &=&
\sum_{k=1}^{N_F} F_2^{\rm VFNS, e_k^2}(x,Q^2)
+ \sum_{Q=c,b}F_2^{\rm VFNS, e_Q^2}(x,Q^2),
\end{eqnarray}
where
\begin{eqnarray}
\label{eq:massl1}
F_2^{\rm VFNS, e_k^2}(x,Q^2) &=& e_k^2 \Biggl[
f_{k+\bar{k}}(x,Q^2,N_F+1) \otimes
C_2^{\rm NS}\left(x,\frac{Q^2}{\mu^2},N_F+1\right)
+ \Sigma(x,\mu^2,N_F+1) 
\nonumber\\ &&
\otimes
\tilde{C}_{2,q}^{\rm PS}\left(x,\frac{Q^2}{\mu^2},N_F+1\right)
+ G(x,\mu^2,N_F+1) \otimes
\tilde{C}_{2,g}^{\rm S}\left(x,\frac{Q^2}{\mu^2},N_F+1\right)\Biggr],
\nonumber\\
\\
\label{eq:massiv1}
F_2^{\rm VFNS, e_Q^2}(x,Q^2) &=& e_Q^2 \Biggl[
f_{Q+\bar{Q}}(x,Q^2,N_F+1) \otimes
C_2^{\rm NS}\left(x,\frac{Q^2}{\mu^2},N_F+1\right)
+ \Sigma(x,\mu^2,N_F+1) 
\nonumber\\ &&
\otimes
\tilde{C}_{2,q}^{\rm PS}\left(x,\frac{Q^2}{\mu^2},N_F+1\right)
+ G(x,\mu^2,N_F+1) \otimes
\tilde{C}_{2,g}^{\rm S}\left(x,\frac{Q^2}{\mu^2},N_F+1\right)\Biggr].
\nonumber
\\
\end{eqnarray}
Here the functions $C_{2,i}^{(k)}$ denote the massless Wilson coefficient 
\cite{Vermaseren:2005qc,Blumlein:2022gpp}, with $\tilde{f}(N_F) = f(N_F)/N_F$.
Applying the renormalization group operator to $F_2(x,Q^2)$
\begin{eqnarray}
\left[\mu \frac{\partial}{\partial \mu}
+ \beta(g,N_F) \frac{\partial}{\partial g}\right] F_2(x,Q^2) =
\frac{d}{dt} F(x,Q^2) = 0,~~\text{with}~~t =
- 2 \ln\left(\frac{\mu_0^2}{\mu^2}\right)
\end{eqnarray}
results in
\begin{eqnarray}
\delta F_2(x,Q^2) =
F_2^{\rm total}(x,Q^2) - F_2^{\rm VFNS}(x,Q^2) = O(a_s^4),
\end{eqnarray}
where $a_s = \alpha_s/(4\pi)$ denotes the strong coupling constant.
Thus the representation in the VFNS agrees to the one in the fixed flavor number scheme
in the range of large values of $Q^2$ at three-loop accuracy.

\section{Single-mass heavy-quark effects on parton distribution functions}
\label{sec:4}

\vspace*{1mm}
\noindent
We have designed numerical $x$-space codes for all unpolarized and polarized single-mass OMEs
to three-loop order in {\tt C++} and {\tt Fortran} in Ref.~\cite{Ablinger:2025joi}. In addition, we 
have released $x$-space codes for the massless and asymptotic massive Wilson coefficients to 
three-loop order in Refs.~\cite{Ablinger:2025awb,Ablinger:2025xqr}. The codes are very fast 
and precise and consist of analytic small- and large-$x$ expansions and cubic splines over 
fine-grided data arrays for the remainder part, leaving $a_s$, $\ln(Q^2/\mu^2), 
\ln(m^2_Q/\mu^2), 
N_F$, and charge weights as free parameters. They can be efficiently used in QCD-fitting codes. 

\begin{center}
\begin{figure}[H]
\centering
\includegraphics[width=\textwidth]{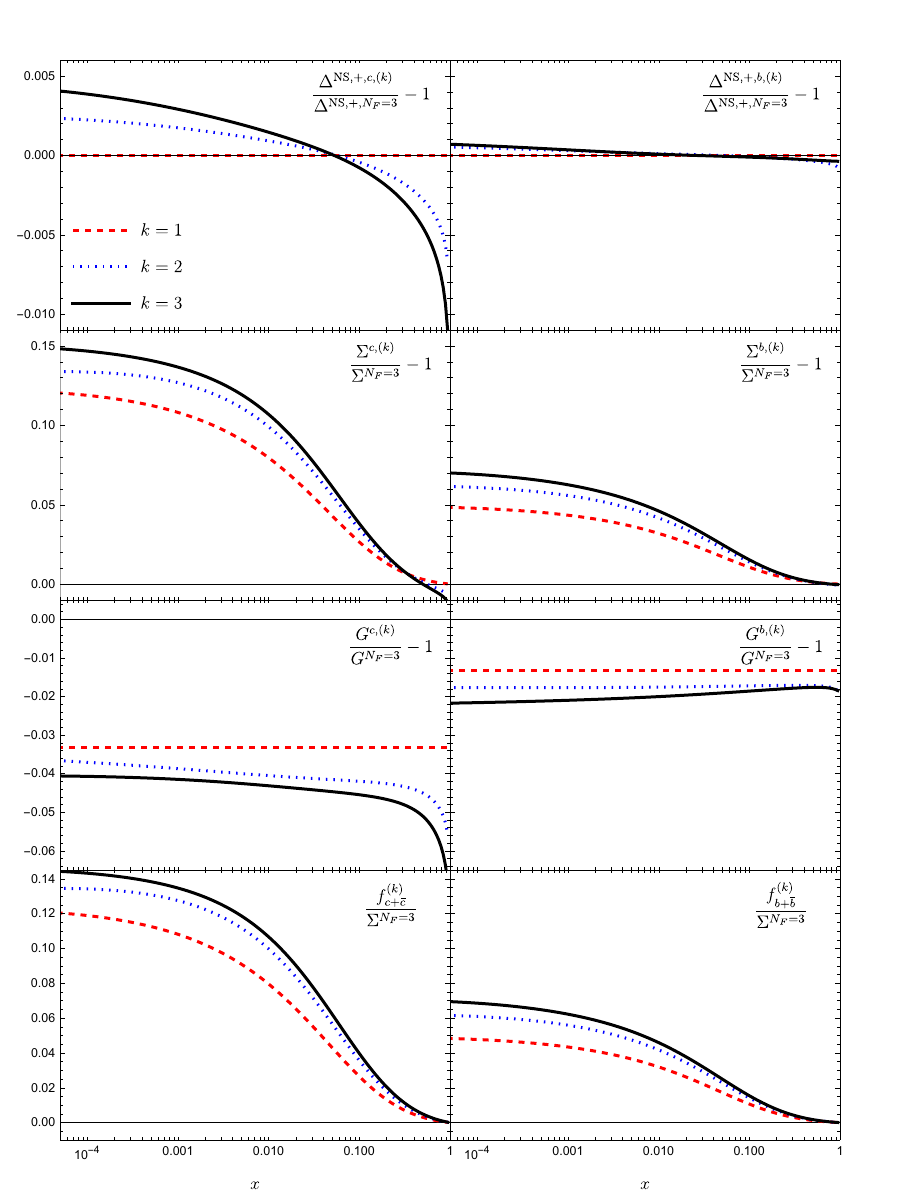}
\caption[]{ \sf
    \label{FIG1}
Left panels: relative heavy-quark contributions due to charm to $\Delta^{\rm NS,+},
\Sigma, G$ and $f_{c+\bar{c}}$. Right panels: relative heavy-quark contributions due to bottom to 
$\Delta^{\rm NS,+}, \Sigma, G$ and $f_{b+\bar{b}}$, both at $Q^2 = 100~\GeV^2$.
The massive OMEs have been
  truncated to the perturbative order $O(a_s^k)$. The denominators of the ratios are
  $\Sigma^{\rm N_F=3}$ or $G^{\rm N_F=3}$, respectively.
Dashed lines: $O(a_s)$. Dotted lines: contributions to $O(a_s^2)$. Full lines:
contributions to $O(a_s^3)$; from Ref.~\cite{Ablinger:2025joi}.}
\end{figure}
\end{center}

In Figure~\ref{FIG1} we illustrate the pure single-mass heavy-flavor contributions
divided by the massless contributions. We also show the degree of convergence from
the $O(a_s)$ to $O(a_s^3)$ in the unpolarized case for the non-singlet, singlet and gluon 
distributions, as well as for the heavy-quark distribution functions.

\begin{figure}[H]\centering
\includegraphics[width=.65\textwidth]{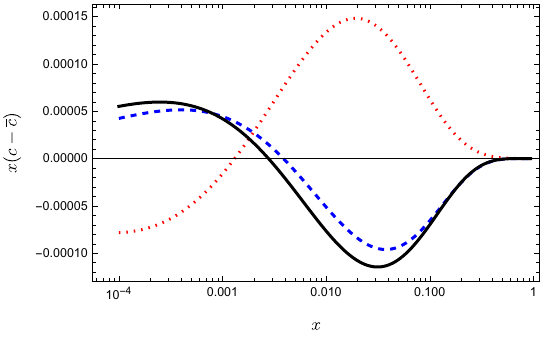}
\caption[]{\sf
The unpolarized  distributions $x[c(x,Q^2) - \overline{c}(x,Q^2)]$; 
from Ref.~\cite{Behring:2025avs}. \label{FIG2}}
\end{figure}

Starting from three-loop order, there are also $f_Q - f_{\bar{Q}}$ distribution functions. These
distributions turn out to be rather small, cf.~Ref.~\cite{Behring:2025avs}, and are shown
in Figure~\ref{FIG2}.\footnote{In the polarized case the associated three-loop anomalous
dimension has been newly calculated by three different methods for the first time
in Ref.~\cite{Behring:2025avs} and independently in Ref.~\cite{Zhu:2025gts}.}

\noindent
Numerical results for the single-mass heavy-flavor three-loop Wilson coefficients of the
structure functions $F_L(x,Q^2)$ \cite{Bierenbaum:2009mv,Bierenbaum:2007qe},  
$xF_3(x,Q^2)$ \cite{Behring:2015roa} and $g_1^{\rm NS}(x,Q^2)$ \cite{Behring:2015zaa}, 
as well as for the charged current structure functions $F_L^{W^+-W^-}(x,Q^2)$ and 
$F_2^{W^+-W^-}(x,Q^2)$ \cite{Behring:2016hpa} and the unpolarized two-loop charged current 
structure functions \cite{Blumlein:2014fqa} have been presented before.\footnote{We have 
corrected typographical errors in $H_{3,g}^{W,(2)}(x)$, \cite{Blumlein:2014fqa}, 
Eq.~(C.11), and agree with 
Ref.~\cite{Caola:2026kvl}. Our $N$-space results were confirmed by 
Ref.~\cite{Caola:2026kvl}.}
We attach the 
corresponding codes as ancillary files to this paper.
Here the same representation for the external variables as in 
Ref.~\cite{Ablinger:2025awb} is used. The functions are provided in terms of harmonic polylogarithms 
\cite{Gehrmann:2001pz,Ablinger:2018sat}.\footnote{In the design of the {\tt Fortran} programmes
we used code optimization \cite{Ruijl:2017dtg}.}

\section{Two-mass operator matrix elements}
\label{sec:5}

\begin{figure}
\includegraphics[width=.49\textwidth]{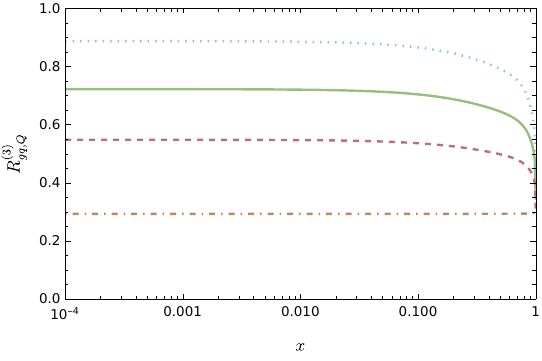}
\includegraphics[width=.49\textwidth]{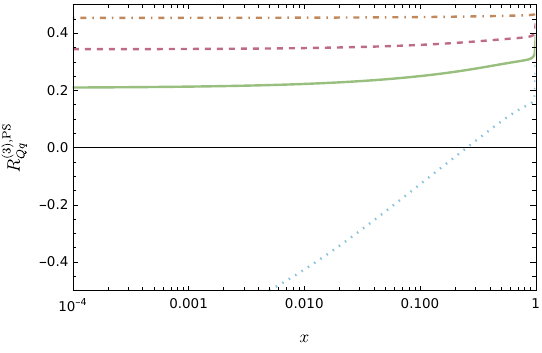}
\caption[]{\sf The contributions of the two-mass OMEs $A_{gq}^{\rm two}$ and 
$A_{Qq}^{\rm two,PS}$ normalized to the complete contributions $O(T_F^2)$
in the unpolarized case.
Dotted lines:      $Q^2 =   30 \GeV^2$ 
Full lines:        $Q^2 =   50 \GeV^2$ 
Dashed  lines:     $Q^2 =  100 \GeV^2$ 
Dash-dotted lines: $Q^2 = 1000\GeV^2$;
from Refs.~\cite{Ablinger:2025xqr,Ablinger:2017xml}. \label{FIG3}}
\end{figure}

\vspace*{1mm}
\noindent
The two-mass operator matrix elements were all calculated to three-loop order in the 
unpolarized 
and polarized cases. 
They consist of factorizable and, from three-loop 
order, also non-factorizable 
contributions. The emergence of two masses complicates the respective integrals and 
functional representation of the final results.\footnote{For the calculation methods we
refer to the original papers and the survey \cite{Blumlein:2018cms}.} 
The respective contributions at 
three-loop order have color factors of $T_F^2 C_{F,A}$.\footnote{In QCD  
the color factors are $T_F = 1/2, C_F = (N_c^2-1)/(2 N_c), C_A = N_c, N_c = 3$.}
The relative impact of the two-mass corrections can be illustrated comparing to the 
complete contributions of $O(T_F^2 C_{F,A})$ in the respective case. We show this
in Figure~\ref{FIG3}.

\noindent
The ratios are of the size $O(1/2)$. Precise numerical codes for the 
two-mass corrections will be released in the same form as in the single-mass case very 
soon.

\section{Conclusions}
\label{sec:6}

\vspace*{1mm}
\noindent
The single- and two-mass VFNS has been promoted from two- to three-loop order both in the
unpolarized and polarized case. The mass-modified parton distribution functions are uniquely 
implied by the massive renormalization group equation. These schemes describe the 
deep-inelastic structure functions in the region $Q^2 \gg m_Q^2$, where power corrections 
can be safely neglected.\footnote{Analytic results for the full phase space were obtained at 
$O(\alpha_s^2)$ in Refs.~\cite{Buza:1995ie,Blumlein:2016xcy,Blumlein:2019qze,Blumlein:2019zux}
and can be used to determine the scale $Q_0^2$ estimating the agreement with the asymptotic
representations at the per-cent level.} Fast and precise numerical implementations were 
released as public codes for the use in data analyses in the massless and single-mass 
case 
\cite{Ablinger:2025joi}. 
This also applies to the massive three-loop Wilson coefficients \cite{Ablinger:2025awb} and the 
massless three-loop Wilson coefficients \cite{Ablinger:2025xqr}. 

The present framework allows for more precise calculations of different processes 
for which the VFNS can be used. This applies in particular to processes at hadron 
colliders. A consistent treatment of the heavy-flavor corrections is of central 
importance for the measurement of the strong coupling constant $a_s$ and the 
heavy-quark masses using high-energy processes at different colliders \cite{dEnterria:2022hzv,
Alekhin:2012vu}, if gluon and flavor-singlet contributions are involved. 

However, for a high-precision measurement
of $a_s$ from the world-data including the future proton and deuteron data taken at the 
EIC \cite{Boer:2011fh}, we recommend a strict 
non-singlet analysis at N$^3$LO, for which already 
now all relevant massless, single- and two-mass corrections are available, 
cf.~Refs.~\cite{Vermaseren:2005qc,Ablinger:2014vwa,Ablinger:2017err,Blumlein:2021lmf,
Blumlein:2022gpp,Gehrmann:2026qbl}, to accomplish a long-standing proposal, 
cf.~Ref.~\cite{Guyot:1988sa}, after about 40 years.

\vspace*{5mm}
\noindent
{\bf Acknowledgments.}
This work was supported by the European Research Council (ERC)
under the European Union's Horizon 2020 research and innovation programme
grant agreement 101019620 (ERC Advanced Grant TOPUP), the UZH Postdoc Grant,
grant no.~[FK-24-115] and by the Austrian Science Fund (FWF) Grant DOI 10.55776/P33530.

{\small

}
\end{document}